# DeFT-AN: Dense Frequency-Time Attentive Network for Multichannel Speech Enhancement

Dongheon Lee, *Student Member, IEEE*, Jung-Woo Choi, *Member, IEEE*

*Abstract*—In this study, we propose a dense frequency-time attentive network (DeFT-AN) for multichannel speech enhancement. DeFT-AN is a mask estimation network that predicts a complex spectral masking pattern for suppressing the noise and reverberation embedded in the short-time Fourier transform (STFT) of an input signal. The proposed mask estimation network incorporates three different types of blocks for aggregating information in the spatial, spectral, and temporal dimensions. It utilizes a spectral transformer with a modified feed-forward network and a temporal conformer with sequential dilated convolutions. The use of dense blocks and transformers dedicated to the three different characteristics of audio signals enables more comprehensive enhancement in noisy and reverberant environments. The remarkable performance of DeFT-AN over state-of-the-art multichannel models is demonstrated based on two popular noisy and reverberant datasets in terms of various metrics for speech quality and intelligibility.

## I. Introduction

Multichannel speech enhancement comprises reconstructing clean speech from noisy multichannel speech. Traditionally, beamforming techniques have been widely adopted for multichannel speech enhancement. Recently, deep learning and deep neural network (DNN)-based methods have gained considerable attention owing to their superior performance. Unlike a single-channel speech enhancement task for spectral and temporal information, multichannel speech enhancements can further improve performance under various conditions by exploiting the spatial information between multichannel data.

Deep learning-based speech enhancement or separation approaches can be categorized into two types depending on their input/output data type [1]. Enhancement techniques estimating spectrogram masks were originally developed for short-time Fourier transform (STFT) data [2, 3]. However, end-to-end methods using time-domain waveforms [4, 5] have been developed to overcome the shortcomings of STFT approaches such as the phase reconstruction issue [5]. In particular, with the discovery of time-domain chunking and dual-path processing [6], methods exploiting short- and long-term temporal dependencies have been proposed [7, 8].

Numerous existing time-domain end-to-end models use the scale-invariant speech-to-distortion ratio (SI-SDR) [9] as a loss function. Moreover, various STFT-based loss functions [10, 11] reflecting time-frequency (TF) characteristics have also been suggested. For example, the phase-constrained magnitude (PCM) loss [11] is defined in terms of magnitudes of real/imaginary (RI) STFT components and has demonstrated an enhanced performance over other loss functions. The STFT can be applied to the output of a time-domain end-to-end model to calculate an STFT-based loss function. However, the direct aggregation and control of the spectral information can be beneficial to optimizing the STFT-based loss [11]. Based on the advantages of spectrograms and solving the phase reconstruction issue in complex domains, STFT-based approaches are being rediscovered. For instance, a recent study reported that spectral information can be more valuable than temporal information [12]. Additionally, other STFT-based approaches [13]–[17] have been proposed for directly learning spectral information. These studies stress that the utilization of spectral information remains an emerging issue in speech enhancement.

In single-channel speech enhancement, models with transformers [18] have been proposed for extracting spectral and temporal information [13, 14]. The transformer has been widely adopted in speech enhancement owing to its parallel processing capabilities and superior performance compared with a recurrent neural network (RNN) [19] or convolutional neural network (CNN). In addition, a conformer architecture [20] proposed for speech recognition is being applied in speech enhancement [21]. These advances in single-channel speech enhancement can be extended to multichannel models for further performance improvements.

To date, representative multichannel speech enhancement models include a deep convolution recurrent network [15] based on a multi-microphone complex spectral mapping for the prediction of RI components, and a triple-path attentive recurrent network (TPARN) [22] utilizing spatial, long-term, and short-term temporal relations. In another approach called the attentive dense convolutional network (ADCN) [16], a U-Net architecture is combined with a temporal attention module. Liu et al. [17] introduced a densely connected recurrent convolutional

This work was supported by the BK21 Four program through National Research Foundation (NRF) funded by the Ministry of Education of Korea, and by the National Research Council of Science & Technology (NST) grant by the Korean government (MSIT) (No. CRC21011) and conducted by the Center for Applied Research in Artificial Intelligence (CARAI) grant funded by DAPA and ADD (UD190031RD). (*Corresponding author: Jung-Woo Choi.*)
Dongheon Lee and Jung-Woo Choi are with the Korea Advanced Institute of Science and Technology (KAIST), Daejeon 34141, South Korea (e-mail: donghen0115@kaist.ac.kr; jwoo@kaist.ac.kr).

neural network (DRC-Net) for combining the advantages of the RNN and CNN. DRC-Net is also based on U-Net, but each layer has a DRC block with a channel-wise BLSTM applied to both the time and frequency dimensions, followed by 2D convolution (Conv) for reducing the increased channel dimensions owing to the dense connections. To summarize, recent models incorporate attention modules to utilize long-term and short-term temporal relations [22] or employ dense blocks to aggregate spatial information without increasing the channel dimensions [16, 17]. The model proposed herein also inherits these advantages owing to the use of dense blocks and transformers. Furthermore, we attempt to combine the advantages of the STFT-based complex spectral masking network and resolve the parallel processing problem of the RNN.

In this study, we propose a novel complex spectral masking network called the dense frequency-time attentive network (DeFT-AN). This model has a series of sub-blocks consisting of a dense block for aggregating spatial information and two transformer blocks for handling spectral and temporal information, respectively. To enable more comprehensive analysis and synthesis of the spectral information, we introduce an F-transformer for focusing on the spectral information, followed by a T-conformer designed to realize the parallelizable architecture without losing the local information or receptive field in time. This is possible by combining a temporal convolutional network (TCN) [23] structure with the attention module. Finally, we demonstrate the performance of DeFT-AN relative to other state-of-the-art approaches based on training and testing over two noisy reverberant datasets.

## II. MODEL ARCHITECTURE

### A. Problem Statement

A noisy reverberant speech signal measured by the $m$ th microphone of a microphone array ($m = 1, \cdots, M$) can be expressed as

$$y_m[n] = x_m[n] + z_m[n] = s_m[n] + r_m[n] + z_m[n], \quad (1)$$

where $n = 0, 1, \cdots, N-1$ is the time index, the signals $s_m$ and $r_m$ denote direct and reverberant speeches without noises, respectively, and $z_m$ is the noise signal embedded in the measured signal. The goal of multichannel speech enhancement is to estimate the clean speech $s_m$ at a reference microphone from the measured noisy speech signals $y_m$.

### B. Dense Frequency-Time Attentive Network (DeFT-AN)

The overall architecture of DeFT-AN is presented in Fig. 1. The multichannel input $\mathbf{y} = [\mathbf{y}_1, \cdots, \mathbf{y}_M]^T \in \mathbb{R}^{M \times N}$ for $\mathbf{y}_m = [y_m[0], \cdots, y_m[N-1]]^T$ is converted into a complex spectrogram tensor $\mathbf{Y} \in \mathbb{R}^{2M \times F \times T}$ by stacking RI components of STFT, where $F$ and $T$ are the total number of frequency bines and time frames, respectively.
The multichannel STFT $\mathbf{Y}$ enters a mask estimation network comprising a series of dense frequency-time attention (DeFT-

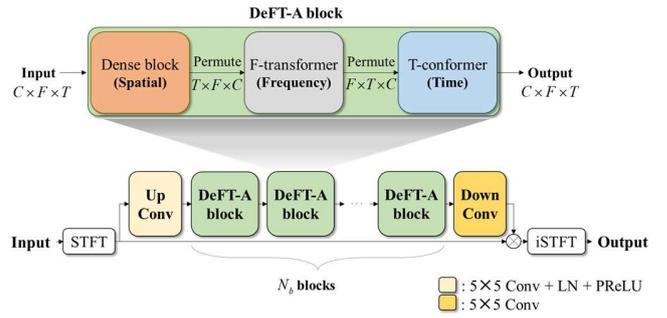

Fig. 1. Overview of the DeFT-AN and DeFT-A block

A) and Up- and Down-Conv blocks. In the first Up-Conv block, the space dimension (microphone channel dimension) is increased from $M$ to $C$ using the 2D Conv operation, followed by layer normalization (LN) and a parametric rectified linear unit (PReLU) activation function [24].

Next, the model extracts and learns the spatial, spectral, and temporal features using a series of DeFT-A blocks. The detailed diagram of the DeFT-A block is presented in Fig. 1. The DeFT-A block incorporates three sub-blocks specialized for learning spatial, spectral, and temporal domain information, respectively. In particular, the spatial information is aggregated primarily in terms of a dense block. The spectral information is processed primarily by the F-transformer, whereas the temporal information is handled by the T-conformer.

The dense block shown in Fig. 2(a) represents a stack of 2D Conv (3×3), LN, and PReLU blocks, and it inherits the structure used in the ADCN [16]. The (channel, width, height) dimensions of the 2D Conv are allocated to the ($C$, $F$, $T$) dimensions of the input. The output from each 2D Conv layer is concatenated with the previous input features. This process is repeated $N_d$ times, therefore, the input to the last 2D Conv layer has a channel dimension size of $N_d C$. The last layer compresses the channel dimension size to $C$.

Next, the dimensions of the dense block output are permuted to ($T$, $F$, $C$). The permuted signal is fed into the F-transformer block, which consists of a spectral attention module and modified feed-forward (FFW) layer (Fig. 2(b)). In the attention module, the time, frequency, and channel dimensions are allocated to the batch, sequence, and feature dimensions, respectively. Herein, a multi-head self-attention (MHSA) module with four heads is used as an attention module that obtains a query, key, and value by using linear projection [18]. The attention output processed by the dropout is added to the attention input and normalized by the LN and fed into the FFW layer. The FFW layer increases the channel dimension size using the 1×1 convolution layer and applies the LN and a Gaussian error linear unit (GELU) [25] activation function. Subsequently, the channel dimension size is reduced in the 1×1 Conv layer connected to the LN and dropout. Finally, the input and output of the FFW layer are summed and normalized by the LN.

Next, the output of the F-transformer block is processed by the T-conformer block depicted in Fig. 2(c). The T-conformer block is designed to focus on the temporal information while employing a sufficient size for the receptive field. This is possible by adopting the TCN structure [5]. First, the frequency,

*B. Experimental Settings*

We extracted the complex spectra of the four microphone signals using an STFT with a rectangular window of 32 ms length (512 samples) and 75% overlap. The RI components of complex spectra with dimensions $F = 257$ were used. We set the total number of DeFT-A blocks as $N_b = 4$, the channel dimension length of the Up-Conv block as $C = 64$, the number of dense blocks as $N_d = 5$, and the number of convolution layers in the SDC as $N_c = 3$ for each DeFT-A block. The kernel size of the DD-Conv in the SDC is 3. All models were trained for 100 epochs with the PCM loss [11] using the Adam optimizer [30] with batch size 1, and the learning rate was fixed at 0.0004. All models discussed in this work are the non-causal system.

All models were evaluated and compared in terms of three objective measures: the SI-SDR [9], perceptual assessment of speech quality (PESQ) [31], and short-time objective intelligibility (STOI) [32]. The STOI scores are presented on a percentage scale. Along with these measures, parameter sizes and computational complexities (MAC/s) were also evaluated.

*C. Parameter Studies*

Parameter studies were conducted using the WSJCAM0 dataset to verify the performance improvement from each block or module. Except for the last experiment, parameters were varied from the best-performance model using 50% overlap. Considered parameter variations and their results are as follows.

*1) Ablation of each sub-block (SUB)*

One of the Dense, T-conformer, and F-conformer blocks was ablated to investigate the effect of each sub-block. The result summarized in Table I indicates that omitting one of these blocks brings significant performance degradation. Removing the Dense block was least notable than the other two ablations, but still, induces the SI-SDR decrease of 4.3 dB compared to the model without ablation.

*2) Subblock order (Order)*

The arranging order of Dense, F-Transformer, and T-Conformer blocks was changed to determine the optimal order of the sub-blocks. The performance was highest with the Dense-Frequency-Time block arrangement, indicating the importance of processing spatial information first. Nevertheless, the impact of sub-block order was subtle. In other parameter studies, the best-performance model was used.

*3) Number of DeFT-A blocks ($N_b$)*

The performance variation with respect to the number of DeFT-A blocks (2, 3, 4) was inspected. All evaluation metrics were improved by increasing the number of DeFT-A blocks. With the repetition of the information aggregation through space, frequency, and time, a more accurate mask could be estimated. However, the performance enhancement was accompanied by doubling parameter size and computational complexity.

*4) Dense frequency-time transformer (DeFT-T)*

The performance of the model using SDC in the T-conformer was compared with those using RNN and GRU. The models using RNN and GRU showed degraded performance, which implies that the benefit of using SDC in extracting temporal features [5] also applies to the FFW of the T-conformer.

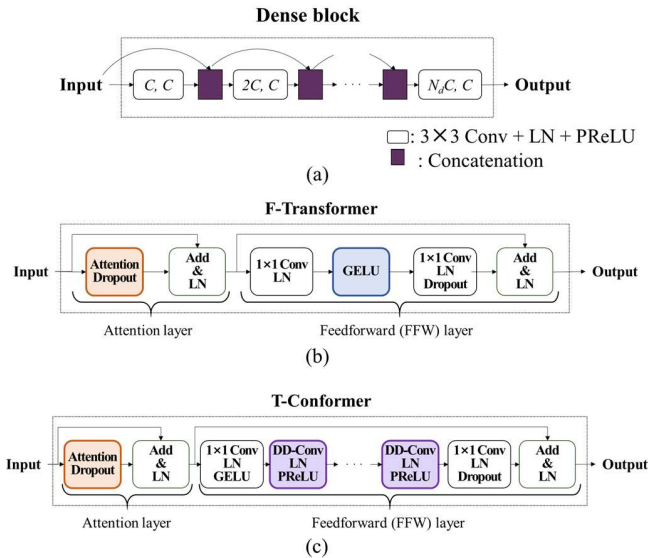

Fig. 2. Diagrams of (a) Dense block, (b) F-Transformer, and (c) T-conformer blocks in DeFT-AN time, and channel dimensions are permuted to the batch, sequence, and feature dimensions of the attention, respectively. The attention part is similar to that of the F-transformer block but connected to the position-wise FFW network [7, 26] to consider temporal context information without positional encoding. In contrast to the conventional position-wise FFW network using RNN or gated recurrent unit (GRU) layers, we utilize a sequential dilated convolutions (SDC) layer consisting of $N_c$ dilated convolution layers with increasing dilation factors. This is used for the parallelization of the temporal convolutions without sacrificing the size of the receptive field. Each dilated convolution layer is given by the combination of the 1D depthwise dilated convolution (DD-Conv), LN, and PReLU. The rest of the architecture is the same as that of the F-transformer block.

## III. EXPERIMENTS

*A. Datasets*

We used spatialized WSJCAM0 [15] and spatialized deep noise suppression (DNS) challenge [22] datasets for the experiment. These two datasets consist of multichannel recordings measured by four microphones ($M = 4$) arranged on a circle with a 10-cm radius. The first dataset was generated by spatializing the speech of the WSJCAM0 dataset [27] using simulated room impulse responses (RIRs) and then adding the noise from the REVERB challenge [28]. The reverberation time (RT60) of RIRs varied within [0.2, 1.3] seconds, and the signal-to-noise ratio (SNR) ranged from 5 dB to 25 dB. A detailed algorithm for creating the spatialized WSJCAM0 dataset is given in [15]. The second dataset was generated by spatializing the speech and noises in the DNS Challenge 2020 corpus [29]. The ranges of RT60 and SNR were [0.2, 1.2] seconds and [−10, 10] dB, respectively. The rooms used for the training and testing sets were all different. The algorithm and parameters for creating the spatialized DNS challenge dataset are described in [22]. All utterances were resampled at 16 kHz, and 4-s-long utterances were randomly extracted during training.

TABLE I
EXPERIMENTAL RESULTS FROM PARAMETER STUDIES

| Parameter | Variations | SI-SDR | PESQ | STOI | Param. Size | MAC/s |
|---|---|---|---|---|---|---|
| SUB | without D | 10.0 | 3.35 | 96.1 | 0.44 M | 10.9 G |
|  | without F | 2.4 | 2.23 | 81.3 | 2.5 M | 43.6 G |
|  | without T | 5.6 | 3.22 | 94.2 | 2.4 M | 41.0 G |
|  | Proposed | **14.3** | **3.52** | **97.4** | 2.7 M | 47.8 G |
| Order | T-F-D | 14.0 | 3.47 | 97.3 | 2.7 M | 47.8 G |
|  | F-T-D | 13.1 | 3.40 | 96.9 | 2.7 M | 47.8 G |
|  | D-F-T | **14.3** | **3.52** | **97.4** | 2.7 M | 47.8 G |
| $N_b$ | 2 | 12.0 | 3.30 | 96.7 | 1.3 M | 26.0 G |
|  | 3 | 13.2 | 3.43 | 97.1 | 2.0 M | 36.9 G |
|  | 4 | **14.3** | **3.52** | **97.4** | 2.7 M | 47.8 G |
| DeFT-T | RNN | 13.2 | 3.39 | 97.0 | 2.9 M | 52.2 G |
|  | GRU | 13.8 | 3.43 | 97.1 | 3.6 M | 64.6 G |
|  | SDC | **14.3** | **3.52** | **97.4** | 2.7 M | 47.8 G |
| Activation | PReLU → ReLU | 13.8 | 3.46 | 97.0 | 2.7 M | 47.8 G |
|  | GELU → ReLU | 13.6 | 3.46 | 96.9 | 2.7 M | 47.8 G |
|  | Proposed | **14.3** | **3.52** | **97.4** | 2.7 M | 47.8 G |
| Overlap | 50% | 14.3 | 3.52 | 97.4 | 2.7 M | 47.8 G |
|  | 75% | **15.7** | **3.63** | **98.1** | 2.7 M | 95.6 G |

TABLE II
PERFORMANCE COMPARISONS WITH BASELINE MODELS

| Dataset | WSJCAM0 | | | | DNS challenge | | | Param. Size | MAC/s |
|---|---|---|---|---|---|---|---|---|---|
| Metrics | SI-SDR | PESQ | STOI | ESTOI | SI-SDR | PESQ | STOI | | |
| Unprocessed | −4.0<br>−3.8[a]<br>−4.0[c] | 1.93<br>1.93[a]<br>1.93[c] | 70.7<br>70.9[b] | 48.6<br><br>48.7[c] | −7.8<br>−7.6[b,d] | 1.38<br>1.38[b,d] | 59.3<br>63.8[b,d] | - | - |
| CA DU-Net [35] | 10.7[d] | 2.88[d] | 96.0[d] | - | 3.5[d] | 1.99[d] | 83.3[d] | 26.0 M | - |
| FasNet TAC [36] | 8.2[d] | 2.93[d] | 94.7[d] | - | 4.7[d] | 2.26[d] | 86.5[d] | 2.9 M | 55.3 G |
| DCRN [15] | 9.4[d] | 3.31[d] | 96.5[d] | - | 4.6[d] | 2.57[d] | 90.1[d] | 18.0 M | 14.4 G |
| TPARN [22] | 10.4[d] | 3.43[d] | 96.9[d] | - | 8.4[d] | 2.75[d] | 91.9[d] | 3.2 M | 71.8 G |
| ADCN [16] | 12.0[b] | 3.42[b] | 97.3[b] | - | 7.8[b] | 2.84[b] | 92.3[b] | 9.3 M | 72.8 G |
| DRC-Net [17] | 12.1<br>12.6[c] | 3.57<br>3.61[c] | 97.5 | 93.2<br>93.5[c] | 6.4 | 2.79 | 89.1 | 2.6 M | 13.9 G |
| **DeFT-AN** | **15.7** | **3.63** | **98.1** | **95.3** | **9.9** | **3.01** | **92.4** | 2.7 M | 95.6 G |
| 2 stage [16] ADCN-TPARN | 13.8[b] | 3.64[b] | 98.0[b] | - | 10.0[b] | 2.99[b] | 93.7[b] | 40.4 M | 360.0 G |
| **2 stage DeFT-AN** | **15.8** | **3.73** | **98.2** | **95.4** | **10.7** | **3.10** | 93.0 | 5.4 M | 191.2 G |

a, b, c, d: results reported in [15], [16], [17], [22] respectively. Numbers without superscripts represent the results reproduced in this work.

TABLE III
EXPERIMENTAL RESULTS OF DEREVERBERATION PERFORMANCE

|  | unprocessed | DCRN | DRC-Net | DeFT-AN |
|---|---|---|---|---|
| SRMR | 4.63 | 6.76 | 7.20 | **7.35** |

*5) Types of activation functions (Activation)*

The role of PReLU and GELU activation functions was examined by replacing them with ReLU [33]. Both modifications brought consistent performance decreases, which again demonstrates the advantages of PReLU for temporal convolution [5] and GELU for FFW layers.

*6) Overlap of STFT window (Overlap)*

The previous study [34] reported that a reduced hop size in STFT increases performance. In this experiment, model performances for 50% and 75% overlaps were compared. The performance was further increased with a 75% overlap. With increasing overlap, the total number of time frames of STFT is doubled, thus allowing a smoother transition from frame to frame at the expense of increased training time.

### D. Comparison to the baseline models

Next, the performance of the proposed model was compared with those of recently proposed multichannel speech enhancement models and presented in Table II. The data reported in [16, 17], and [22], including ESTOI [37], were used for the performance of the baseline models. However, some discrepancies can exist with different realizations of the dataset, so we also provide the performance of the reproduced baseline models and unprocessed data when available.

Both experiments with the spatialized WSJCAM0 dataset and the spatialized DNS challenge dataset show that the DeFT-AN outperforms the state-of-the-art models by a large margin in terms of most evaluation measures. Notably, the proposed method exhibits highly improved SI-SDR and PESQ values relative to those of the baseline models.

The proposed method was also compared with the two-stage approach [16] using ADCN for the first stage and TPARN for the second stage (ADCN-TPARN). The single-stage performance of the proposed model is similar to that of the two-stage approach, but the 2-stage setting of the proposed model again outperforms ADCN-TPARN. One exceptional case is the STOI result from the DNS challenge dataset, which is suspected to originate from the discrepancy in STOI value of the unprocessed data used in this work (59.3%) and that used for the reported results (63.8%) [22]. In view of the initial reported penalty of 4.5% owing to the dataset difference, the STOI performance of the proposed method is within a reasonable range.

Another advantage of the proposed model is its small parameter size (2.7 M) comparable to the DRC-Net. For the computational complexity, the attention-based models (TPARN, ADCN, DeFT-AN) exhibit higher MAC/s than others (DCRN, DRC-Net). The complexity of our model is the highest among the single-stage models but much lower than the two-stage model (ADCN-TPARN) in both single- and two-stage settings.

Lastly, the dereverberation performance of the proposed model was separately evaluated by using noiseless test data simulated from unseen reverberant environments of the spatialized WSJCAM0 dataset. Since our objective is to evaluate the dereverberation performance of speech enhancement models performing both denoising and dereverberation, the speech-to-reverberation modulation energy ratio (SRMR) [38] was calculated for the baseline models. The results of Table III demonstrate that the DeFT-AN surpasses other dereverberation models, and hence, can robustly reduce reverberation and enhance speech in the unseen environment.

## IV. CONCLUSION

In this study, a dense frequency-time attentive network (DeFT-AN) was proposed for multichannel speech enhancement. DeFT-AN is a combination of a dense block responsible for analyzing spatial information and two transformer blocks dedicated to spectral and temporal relations, respectively. The proposed method is parallelizable and has an adequate receptive field by using the SDC. Ablation studies revealed that a series connection of DeFT-A blocks with sufficient overlaps between frames could improve the enhancement performance. From training and testing with the spatialized WSJCAM0 and DNS datasets, we demonstrated that the proposed model can achieve speech enhancement in unseen noisy reverberant environments.

REFERENCES

[1] D. L. Wang and J. Chen, "Supervised speech separation based on deep learning: An overview," *IEEE/ACM Trans. Audio, Speech, Lang. Process.*, vol. 26, no. 10, pp. 1702–1726, May. 2018.

[2] D. L. Wang, "On ideal binary mask as the computational goal of auditory scene analysis," *Speech Separation by Humans and Machines*. Boston, MA: Springer, 2005.

[3] A. Narayanan and D. L. Wang, "Ideal ratio mask estimation using deep neural networks for robust speech recognition," in *Proc. IEEE Int. Conf. Acoust., Speech Signal Process.*, 2013, pp. 7092–7096.

[4] S. Venkataramani and J. Casebeer, "End-to-end source separation with adaptive front-ends," in *Proc. 52nd Asilomar Conference on Signals, Systems, and Computers*, 2018, pp. 684–688.

[5] Y. Luo and N. Mesgarani, "Conv-tasnet: Surpassing ideal time-frequency magnitude masking for speech separation," *IEEE/ACM Trans. Audio Speech, Lang. Process.*, vol. 27, no. 8, pp. 1256–1266, May. 2019.

[6] Y. Luo, Z. Chen and T. Yoshioka, "Dual-path RNN: Efficient long sequence modeling for time-domain single-channel speech separation," in *Proc. IEEE Int. Conf. Acoust., Speech Signal Process.*, 2020, pp. 46–50.

[7] J. Chen, Q. Mao, and D. Liu, "Dual-path transformer network: Direct context-aware modeling for end-to-end monaural speech separation," in *Proc. Interspeech*, 2020, pp. 2642–2646.

[8] C. Subakan, M. Ravanelli, S. Cornell, M. Bronzi, and J. Zhong, "Attention is all you need in speech separation," in *Proc. IEEE Int. Conf. Acoust., Speech Signal Process.*, 2021, pp. 2125.

[9] J. Le Roux, S. Wisdom, H. Erdogan, and J. R. Hershey, "SDR– half-baked or well done?," in *Proc. IEEE Int. Conf Acoust., Speech Signal Process.*, 2019, pp. 626–630.

[10] A. Pandey and D. L. Wang, "A new framework for CNN-Based speech enhancement in the time domain," *IEEE/ACM Trans. Audio, Speech, Lang. Process.*, vol. 27, no. 7, pp. 1179–1188, Jul. 2019.

[11] A. Pandey and D. L. Wang, "Dense CNN with self-attention for time-domain speech enhancement," *IEEE/ACM Trans. Audio, Speech, Lang. Process.*, vol. 29, pp. 1270-1279, March, 2021.

[12] K. Tesch, N-H. Mohrmann and T. Gerkmann, "On the role of spatial, spectral, and temporal processing for DNN-based non-linear multi-channel speech enhancement," in *Proc. Interspeech*, 2022, pp. 2098–2912.

[13] L. Yang, W. Liu, and W. Wang, "TFPSNet: Time-frequency domain path scanning network for speech separation," in *Proc. IEEE Int. Conf. Acoust., Speech Signal Process.*, 2022, pp. 6842–6846.

[14] Z. Q. Wang, S. Cornell, S. Choi, Y. Lee, B. Y. Kim, and S. Watanabe, "TF-GridNet: Making Time-Frequency domain models great again for monaural speaker separation," 2022, *arXiv:2209.03952*.

[15] Z. Q. Wang and D. L. Wang, "Multi-microphone complex spectral mapping for speech dereverberation," in *Proc. IEEE Int. Conf. Acoust., Speech Signal Process.*, 2020, pp. 486–490.

[16] A. Pandey, B. Xu, A. Kumar, J. Donley, P. Calamia, and D. L. Wang, "Multichannel speech enhancement without beamforming," in *Proc. IEEE Int. Conf. Acoust., Speech Signal Process.*, 2022, pp. 6502–6506.

[17] J. Liu and X. Zhang, "DRC-NET: Densely connected recurrent convolutional neural network for speech dereverberation," in *Proc. IEEE Int. Conf. Acoust., Speech Signal Process.*, 2022, pp. 166–170.

[18] A. Vaswani *et al.*, "'Attention is all you need,' in *Proc. Advances in Neural Inf. Process. Syst.*, vol. 30, 2017, pp. 5998–6008.

[19] J. L. Elman, "Finding structure in time," *Cogn. Sci.*, vol. 14, no. 2, pp. 179–211, Mar. 1990, doi: 10.1207/s15516709cog1402_1.

[20] A. Gulati *et al.*, "Conformer: Convolution-augmented transformer for speech recognition," in *Proc. Interspeech*, 2020, pp. 5036–5040.

[21] Y. Koizumi *et al.*, "DF-Conformer: Integrated architecture of Conv-TasNet and Conformer using linear complexity self-attention for speech enhancement," in *IEEE Workshop Appl. Signal Process. Audio Acoust.*, pp. 161–165, 2021.

[22] A. Pandey, B. Xu, A. Kumar, J. Donley, P. Calamia, and D. L. Wang, "TPARN: Triple-path attentive recurrent network for time-domain multichannel speech enhancement," in *Proc. IEEE Int. Conf. Acoust., Speech Signal Process.*, 2022, pp. 6497–6501.

[23] C. Lea, M. D. Flynn, R. Vidal, A. Reiter, and G. D. Hager, "Temporal convolutional networks for action segmentation and detection," in *Proc. IEEE Conf. Comput. Vis. Pattern Recognit.*, 2017, pp. 156–165.

[24] K. He, X. Zhang, S. Ren, and J. Sun, "Delving deep into rectifiers: Surpassing human-level performance on imagenet classification," in *Proc. of the IEEE Int. Conf. on Comput. Vis.*, 2015, pp. 1026–1034.

[25] D. Hendrycks and K. Gimpel, "Gaussian Error Linear Units (gelus)," 2016, *arXiv:1606.08415*.

[26] K. Wang, B. He and W. P. Zhu, "TSTNN: Two-stage transformer based neural network for speech enhancement in the time domain," in *Proc. IEEE Int. Conf. Acoust., Speech Signal Process.*, 2021, pp. 7098–7102.

[27] T. Robinson, J. Fransen, D. Pye, J. Foote, and S. Renals, "WSJCAM0: A British English speech corpus for large vocabulary continuous speech recognition," in *Proc. IEEE Int. Conf. Acoust., Speech Signal Process.*, 1995, pp. 81–84.

[28] K. Kinoshita *et al.*, "A summary of the REVERB challenge: State-of-the-art and remaining challenges in reverberant speech processing research," *EURASIP J. Adv. Signal Process.*, vol. 2016, no. 1, pp. 1–19, Jan. 2016, doi: 10.1186/s13634-016-0306-6.

[29] C. K. Reddy *et al.*, "The INTERSPEECH 2020 deep noise suppression challenge," in *Proc. Interspeech*, 2020, pp. 2492–2496.

[30] D. P. Kingma and J. Ba, "Adam: A Method for Stochastic Optimization," in *Proc. Int. Conf. Lear. Represent.*, 2015. [Online]. Available: http://arxiv.org/abs/1412.6980

[31] A. W. Rix, J. G. Beerends, M. P. Hollier, and A. P. Hekstra, "Perceptual evaluation of speech quality (PESQ) – A new method for speech quality assessment of telephone networks and codecs" in *Proc. IEEE Int. Conf. Acoust., Speech Signal Process,* 2001, pp. 749–752.

[32] C. H. Taal, R. C. Hendriks, R. Heusdens, and J. Jensen, "Short-time objective intelligibility measure for time-frequency weighted noisy speech," in *Proc. IEEE Int. Conf. Acoust., Speech Signal Process.*, 2010, pp. 4214–4217.

[33] V. Nair and G. E. Hinton, "Rectified linear units improve restricted Boltzmann machines," in *Proc. International Conference on Machine Learning*, 2010, pp. 807–814.

[34] A. Pandey and D. L. Wang, "On cross-corpus generalization of deep learning based speech enhancement," *IEEE/ACM Trans. Audio, Speech, Lang. Process.*, vol. 28, pp. 2489–2499, Aug. 2020, doi: 10.1109/taslp.2020.3016487.

[35] B. Tolooshams *et al.*, "Channel-attention dense U-Net for multi-channel speech enhancement," in *Proc. IEEE Int. Conf. Acoust., Speech Signal Process.*, 2020, pp. 836–840.

[36] Y. Luo, Z. Chen, N. Mesgarani, and T. Yoshioka, "End-to-end microphone permutation and number invariant multi-channel speech separation," in *Proc. IEEE Int. Conf. Acoust., Speech Signal Process.*, 2020, pp. 6394–6398.

[37] J. Jensen and C. H. Taal, "An algorithm for predicting the intelligibility of speech masked by modulated noise maskers," *IEEE/ACM Trans. Audio, Speech, Lang, Process.*, vol. 24, no. 11, pp. 2009–2022, Nov. 2016, doi: 10.1109/TASLP.2016.2585878.

[38] T.H. Falk, C. Zheng, and W.Y. Chan, "A non-intrusive quality and intelligibility measure of reverberant and dereverberated speech," *IEEE Trans. Audio, Speech, Lang. Process.*, vol. 18, no. 7, pp. 1766 1774, Aug. 2010.